# Giant X-ray circular dichroism in a time-reversal invariant altermagnet

*Jun Okamoto Ru-Pan Wang Yen-Yi Chu Hung-Wei Shiu Amol Singh Hsiao-Yu Huang Chung-Yu Mou Sucitto Teh Horng-Tay Jeng Kai Du Xianghan Xu Sang-Wook Cheong Chao-Hung Du Chien-Te Chen Atsushi Fujimori\* Di-Jing Huang\**

Jun Okamoto
National Synchrotron Radiation Research Center, Hsinchu 30076, Taiwan

Ru-Pan Wang
Department of Physics, University of Hamburg, Luruper Chaussee 149, G610, 22761 Hamburg, Germany

Yen-Yi Chu, Hung-Wei Shiu, Amol Singh, Hsiao-Yu Huang
National Synchrotron Radiation Research Center, Hsinchu 30076, Taiwan

Chung-Yu Mou
Center for Quantum Science and Technology and Department of Physics, National Tsing Hua University, Hsinchu 30013, Taiwan

Sucitto Teh, Horng-Tay Jeng
Department of Physics, National Tsing Hua University, Hsinchu 30013, Taiwan

Kai Du, Xianghan Xu, Sang-Wook Cheong
Rutgers Center for Emergent Materials and Department of Physics and Astronomy, Rutgers University, Piscataway, NJ 08854, USA

Chao-Hung Du
Department of Physics, Tamkang University, Tamsui 251, Taiwan

Chien-Te Chen
National Synchrotron Radiation Research Center, Hsinchu 30076, Taiwan

Atsushi Fujimori
National Synchrotron Radiation Research Center, Hsinchu 30076, Taiwan
Center for Quantum Science and Technology and Department of Physics, National Tsing Hua University, Hsinchu 30013, Taiwan
Department of Physics, University of Tokyo, Bunkyo-Ku, Tokyo 113-0033, Japan
Email Address: fujimori@phys.s.u-tokyo.ac.jp

Di-Jing Huang
National Synchrotron Radiation Research Center, Hsinchu 30076, Taiwan
Department of Physics, National Tsing Hua University, Hsinchu 30013, Taiwan
Department of Electrophysics, National Yang Ming Chiao Tung University, Hsinchu 30093, Taiwan Email Address: djhuang@nsrrc.org.tw

X-ray circular dichroism, arising from the contrast in X-ray absorption between opposite photon helicities, serves as a spectroscopic tool to measure the magnetization of ferromagnetic materials and identify the handedness of chiral crystals. Antiferromagnets with crystallographic chirality typically lack X-ray magnetic circular dichroism because of time-reversal symmetry, yet exhibit weak X-ray natural circular dichroism. Here, we report the observation of giant natural circular dichroism in the Ni $L_3$-edge X-ray absorption of $Ni_3TeO_6$, a polar and chiral antiferromagnet with effective time-reversal symmetry. To unravel this intriguing phenomenon, we propose a phenomenological model that classifies the movement of photons in a chiral crystal within the same symmetry class as that of a magnetic field. The coupling of X-ray polarization with the induced magnetization yields giant X-ray natural circular dichroism, revealing the altermagnetism of $Ni_3TeO_6$. Our findings provide evidence for the interplay between magnetism and crystal chirality in natural optical activity. Additionally, we establish the first example of a new class of magnetic materials exhibiting circular dichroism with time-reversal symmetry.

# 1 Introduction

The interaction of light with matter provides a powerful technique to investigate the breaking of space inversion and time-reversal (T) symmetries. One can observe circular dichroism (CD), i.e., the differential absorption of right- and left-handed circularly polarized (RCP and LCP) light, in a crystal lacking inversion symmetry. The optical activity that measures X-ray circular dichroism (XCD) in a material with T symmetry is termed X-ray *natural* circular dichroism (XNCD),[1–3] conventionally understood to arise from the interference between electric-dipole (E1) and electric-quadrupole (E2) transitions.[4] Natural optical activity is a reciprocal effect, as the absorption of photons moving in one direction is the same as that in the opposite direction.

For a material with broken T symmetry such as a ferromagnet or a ferrimagnet,[5,6] magnetic circular dichroism in X-ray absorption (XMCD) occurs,[7–14] enabling the direct detection of the spin and orbital moments of an ion. While conventional understanding expects vanishing XMCD from antiferromagnetic (AFM) materials, which have no net spin moments, recent results show that XMCD can exist in AFM materials with non-collinear, namely, noncoplanar or chiral spin structures,[15–21] corroborating the prediction based on symmetry analysis with magnetization in relation to broken symmetries.[22] Recently, a class of "altermagnetic" materials,[23–26] where collinear AFM spins are fully canceled leaving no net spin moments, were found to exhibit strong XMCD due to broken T symmetry.[27]

A chiral crystal is non-superimposable on its mirror image and has well-defined handedness. Crystal chirality alone can lead to XNCD unrelated to magnetism;[1,28] however, its magnitude is a few orders of magnitude smaller than the typical XMCD at the $L_{2,3}$ edges of $3d$ transition-metal elements. Recently, a novel kind of spontaneous Hall effect has been reported in collinear antiferromagnets with local crystal chirality.[29] In addition, chiral crystals[30] as well as chiral molecules including DNA on metal surfaces[31,32] produce strong spontaneous spin polarization in charge current, referred to as chirality-induced spin selectivity (CISS).

The present Article presents the observation of the significant enhancement of XNCD by collinear antiferromagnetism. We discovered a new type of surprisingly strong XNCD at the Ni $L_3$ edge of the multiferroic antiferromagnet $Ni_3TeO_6$, whose crystal structure is polar and chiral but the magnetic structure is collinear.[33] Without an external magnetic field, results of X-ray absorption spectroscopy (XAS) showed that XCD changed its sign between domains of opposite crystal chirality. We explain this surprising observation through a new concept involving symmetry analysis[34] for the mechanism of XNCD.

# 2 Results

## 2.1 Magnetic transition of $Ni_3TeO_6$

Nickel tellurite $Ni_3TeO_6$ crystallizes in a modified corundum structure,[35,36] and exhibits large optical rotation and intriguing chiral and polar domains.[37] There are three Ni sites and one Te site in the formula unit. As shown in **Figure 1**a, two kinds of honeycomb layers formed by edge-sharing $NiO_6$ and $TeO_6$ octahedra stack along the *c*-axis, generating the polar crystal structure. Ni$_I$O$_6$ and Ni$_{II}$O$_6$ are ferromagnetically coupled and form one honeycomb layer, and Ni$_{III}$O$_6$ and TeO$_6$ form the other. Ni$_I$O$_6$ and Ni$_{II}$O$_6$ are also connected to Ni$_{III}$O$_6$ on one side of the honeycomb layer through corner- and face-sharing, respectively, while both are connected to Ni$_{III}$O$_6$ on the other side via corner-sharing. In addition, the relative positions of these octahedra in the *ab*-plane revolve along the *c*-axis, showing the handedness of crystal chirality,[37] as illustrated in **Figure 1**b. There exist two enantiomorphic crystal structures in Ni$_3$TeO$_6$, forming the so-called Dauphiné twin, in which right-handed and left-handed chiral domains are neighboring in the crystal.[37] Kerr rotation has been observed for each chiral domain.[37,38] No structural phase transition has been observed from 11 to 300 K, but internal coordinates such as bond lengths and bond angles vary with temperature.[39]

Ni$_3$TeO$_6$ undergoes a magnetic phase transition at $T_N$ = 52-62 K from a paramagnetic state to a commensurate collinear AFM state with Ising anisotropy as shown in **Figure 1**b.[33,40,41] Momentum scans of resonant X-ray magnetic scattering (RXMS) at various temperatures shown in **Figures 2**a, b reveal that the intensity of the (0, 0, 1.5) peak appears below $T_N$ and steadily increases with decreasing temperature, confirming that the AFM ordering occurs below $T_N$. Results of temperature-dependent RXMS at the Ni $L_3$ edge (**Figures 2**a-c) and magnetic susceptibility (**Figure 2**d) are consistent with the collinear AFM structure of Ni$_3$TeO$_6$. The Néel temperature of the sample studied here was $T_N$ = 62 K. The observed antiferromagnetic transition is marked by a sharp cusp in the parallel component $\chi_\parallel$ of the magnetic susceptibility. These results also agree with the results of X-ray linear dichroism (XLD) shown in Supporting Information **Figure S7** and temperature-dependent XLD plotted in **Figure 2**c.

## 2.2 Ni *L*-edge XAS

The dominant electronic configuration of the Ni$^{2+}$ ion in oxides is typically $d^8$, into which the O 2*p*-to-Ni 3*d* charge-transfer configuration $d^9\underline{L}$ is hybridized as a minor (~ 20%) component.[42] Usually, *p*-to-*d* charge-transfer is considered within a NiO$_6$ octahedron, but charge-transfer between neighboring NiO$_6$ octahedra also occurs and increases the $d^9\underline{L}$ component, if the two octahedra share corner oxygen, and the two Ni spins are aligned antiferromagnetically.[43] This is the case for the neighboring Ni$_I$O$_6$/Ni$_{II}$O$_6$ and Ni$_{III}$O$_6$ octahedra coupled via AFM exchange ($J_4$/$J_3$) in the collinear antiferromagnetic Ni$_3$TeO$_6$ (See **Figure 1**b for spin structure and Supporting Information **Figure S5** for exchange interactions between Ni ion.)[44]

Ni$_3$TeO$_6$ is a remarkably unique material, exhibiting both chirality and polarity. Owing to its polar *R*3 structure, polarity aligns with chirality, making it impossible to observe different polarities within the same chiral domain concurrently. We therefore studied XCD of two chiral domains by performing Ni $L_3$-edge XAS measurements using circularly polarized X-rays focused on each chiral domain. **Figures 3**a,b plot the average of Ni $L_3$-edge XAS spectra taken with RCP and LCP X-rays impinging on positions marked A and B, respectively, in a 90×90-$\mu$m$^2$ image shown in **Figure 4**a. This area contains two different structurally chiral domains. One can see that, around $T \sim T_N$, the peak intensity of XAS is enhanced and shifted towards higher photon energies by ~ 0.3 eV relative to the peak position 853 eV denoted as $\omega_{RT}$ at room temperature. As explained below, this suggests an increase in the number of *d* holes on the Ni ions. When the temperature increases towards $T_N$, the AFM correlation between the neighboring ferromagnetic double layers is weakened, and the *p*-to-*d* charge transfer is reduced. This increases the weight of the $d^8$ configuration relative to the charge-transferred $d^9\underline{L}$ configuration and hence the number of *d* holes on Ni, resulting in the shift of the XAS peak towards higher photon energies and the increase of the XAS intensity simultaneously.

## 2.3 Emergence of giant XCD

Surprisingly, we observed significant XCD at each domain for temperatures below ~ $T_N$, as shown in **Figure 4**b, despite the absence of an external magnetic field nor remanent magnetization. See the magnetization curves in Supporting Information **Figure S3**. The measured dichroic spectra, i.e., the RCP-LCP difference spectra, show a maximum magnitude of XCD about 20% of Ni $L_3$-edge XAS peak, much larger than the previously reported one of 2%,[39] and exhibit a non-linear dependence on the X-ray intensity (See **Figure S4** plotted in Supporting Information.), showing a positive peak at $\omega_{RT}$+ 0.3 eV at position A and a negative peak at $\omega_{RT}$+ 0.7 eV at position B. The dichroic images shown in **Figure 4**c demonstrate a sign change between the domains of opposite chirality. The color contrast between the domains increases with temperature, and becomes strongest around $T_N$ and invisibly weak at room temperature. The XCD intensities at A and B as functions of temperature plotted in **Figure 4**d also show a consistent tendency, particularly the enhancement below ~ $T_N$.

# 3 Discussion

## 3.1 Symmetry analysis

Ni$_3$TeO$_6$ exhibits a chiral lattice and effective time-reversal symmetry in its ground state. The illumination of left-handed polarized X-ray with a wave vector $+\vec{k}$ on the crystallographic left-chiral domain exhibits identical behavior to that with $-\vec{k}$. In other words, the observed XCD is reciprocal for temperatures above and below $T_N$. In contrast, XMCD on ferromagnets is nonreciprocal, i.e., XMCD with $+\vec{k}$ is different from that with $-\vec{k}$. In an AFM state, XMCD signals are typically canceled between the opposite magnetic moments, and only weak XMCD signals can be observed under a strong external magnetic field through the field-induced uniform magnetization (typically of ~ 0.1 $\mu_B$ per Ni atom under a field of ~ 1 T). The magnitude of the observed XCD in Ni$_3$TeO$_6$ is comparable to the XMCD of 3$d$ transition metals but much larger than the usual XNCD resulting from the E1-E2 interference. This phenomenon is a new type of XCD with a mechanism beyond the E1-E2 interference. The XCD observed in the present study appears to be comparable with the XMCD of a magnetic moment of order $\mu_B$ per Ni atom despite the absence of an external magnetic field. Since there are equal numbers of spin-up and spin-down Ni atoms in the collinear AFM state, the XCD signals arising from spin moments must be canceled unless there were a uniform "effective magnetic field" that has not been considered in conventional XMCD mechanisms.

A possible origin of the "effective magnetic field" is the crystal chirality of each domain (**Figure 1**b) because the chirality is uniform in each domain and the dichroic signal changes its sign between the domains. The crystal chirality alone does not break "effective" T symmetry and can lead only to tiny XNCD signals. It, therefore, cannot explain the appreciable dichroic intensity enhancement below $T_N$ as remarked above. From the viewpoint of symmetry, the total Hamiltonian of the whole system including Ni$_3$TeO$_6$ and X-rays should be invariant under the simultaneous transformations of charge conjugation, parity (P), and T.[45] For temperatures above $T_N$, the coupling of circularly polarized X-rays to crystal chirality makes the photon-matter interaction symmetric under P transformation (more precisely, mirror image transformation). That is, an observation of XNCD requires that the sample has a broken PT symmetry.

## 3.2 Origin of the X-ray circular dichroism

The enhancement of XCD in Ni$_3$TeO$_6$ below $T_N$ must arise from the coupling of X-rays to the magnetic order in the presence of crystal chirality. We propose a theory that is consistent with the symmetry analysis[34] to understand this intriguing giant XNCD. Theoretically, to the linear order, the EM fields of X-ray generally induce the coupling of electric current $\vec{j}$ to the vector potential $\vec{A}$ as $-\int d^3 \vec{r} \, \vec{j} \cdot \vec{A}$. As shown in the Supporting Information, the transverse part of the current operator can be written as the magnetization current due to the orbital magnetization, and the EM field of X-ray induces a coupling as $-\int d^3 \vec{r} \, \vec{B} \cdot \vec{M}$. In other words, the movement of X-ray photons within a chiral crystal behaves as a magnetic field. The chiral structure of Ni$_3$TeO$_6$ ensures a non-zero average orbital magnetization. In turn, orbital magnetization will be induced, and through spin-orbit coupling (SOC), spin polarization is further generated. When the wave vector of X-rays is changed from $+\vec{k}$ to $-\vec{k}$, the Poynting vector $\vec{E} \times \vec{B}$ is also reversed and this can be viewed as $\vec{B}$ goes to $-\vec{B}$. Hence, the induced orbital magnetization also changes sign. If $\vec{k}$ is parallel to the axis of a chiral structure that only breaks the mirror symmetry, the orbital current in the chiral structure is reversed with the same magnitude. The reversed orbital magnetization thus has the same magnitude so that the net effect is that $\vec{M}$ goes to $-\vec{M}$. The observed XCD thus is reciprocal, and the reciprocity justifies its name as an XNCD, rather than XMCD.

Microscopically, the Berry curvature[46–48] can be also regarded as an effective magnetic field in $k$ space; it will create circular motion of electrons, resulting in the orbital moment, which can be probed by dipole transition using circularly polarized light. The induced orbital magnetization in the momentum space is reflected as the splitting of the quasi-particle energies and is proportional to the Berry curvature.[47] The Berry curvature vanishes for systems possessing both inversion symmetry and time reversal symmetry.

In the case of a system with time reversal symmetry but lacking inversion symmetry, nonvanishing Berry curvature exists. Formally, we have proved that the XCD cross section of $L$-edge XAS is proportional to the integration of Berry curvature over momentum at a fixed energy (refer to Equation S37 of the Supporting Information). In accordance with the XMCD sum rule,[11] the energy integration of XCD is proportional to the integration of Berry curvature across the entire momentum space, leading to an orbital moment.[47] The connection of XCD to the Berry curvature is important in shaping our understanding of the magnetic properties of materials because we do not necessarily need time-reversal symmetry broken to obtain XCD. Indeed, the non-vanishing Berry curvature in Ni$_3$TeO$_6$ results from inversion-symmetry broken, forming a class of materials different from those based on time-reversal symmetry broken. Our model agrees with the observation of current-induced magnetization in a chiral semiconductor Te, where SOC is strong.[49,50] The incident X-rays can induce chiral phonons and magnons in Ni$_3$TeO$_6$ as it exhibits a chiral structure and SOC is also strong due to the presence of the Te atoms. The angular momenta of the X-rays eventually develop through the creation of such phonons and magnons and continue to enhance the XNCD. In CISS, an electric current injected through chiral molecules[32,51] or crystals,[30] the chiral structure induces spin polarization in a preferred direction and imparts significant spin selectivity to the charge current. Similar spin selectivity has also been observed for chiral molecules even without a charge current.[52] In the insulating Ni$_3$TeO$_6$, a finite heat current due to nonequilibrium phonons and magnons can develop along the temperature gradient[53] and flow instead of a charge current, creating a phonon angular momentum.[54] The spin polarization induced by the incident X-ray in the chiral structure is subsequently enhanced by this "thermal CISS" effect through the excitation of relevant phonons and magnons.

The existence of giant XNCD in $Ni_3TeO_6$ is related to its altermagnetism—magnetic states with antiferromagnetic spins, in which the symmetry allows for typical ferromagnetic behaviors. The observed XNCD exhibits a strong temperature dependency like the XMCD from a ferromagnet across its magnetic phase transition, revealing the characteristic of altermagnetism. $Ni_3TeO_6$ has a symmetry corresponding to the magnetic point group of 31' and is categorized as a type-III altermagnet,[26] that has broken PT symmetry and does not exhibit odd-order anomalous Hall effect. A type-III altermagnet exhibits ferromagnetic behaviors only with external perturbations that conserve PT symmetry. In the present XNCD measurement, the significant enhancement in the XNCD of $Ni_3TeO_6$ across $T_N$ reveals its altermagnetism; the moving X-ray photons serve as the external perturbations, and induce magnetization along the X-ray moving direction. Additionally, the observed giant XNCD appears to reflect the magnetic susceptibility of $Ni_3TeO_6$. The observation XNCD below $T_N$ with a magnitude comparable to that of XMCD suggests an effective coupling between spin and chirality. Although the orbital magnetization is time-reversal odd and it alone can't couple to the structural chirality directly, the combination of X-ray propagation and its induced magnetization is time-reversal even, like structural chirality. RXMS intensity maps depicted in **Figure S2** show images consistent with chiral structure images, supporting the above scheme of effective spin chiral coupling. Therefore, the propagation of incident X-rays in a chiral structure generates an effective magnetic field capable of inducing an orbital magnetic moment and hence spin polarization through SOC. Consequently, the XNCD measurement contains the magnetic susceptibility $\chi$. Temperature-dependent XNCD results, as shown in **Figure 4**d, demonstrate a finite magnetic susceptibility in $Ni_3TeO_6$ above $T_N$. The increase in XNCD near $T_N$ corresponds to the rise in $\chi$ as depicted in **Figure 2**d. Below $T_N$, XNCD encompasses both $\chi_\perp$ and $\chi_\parallel$ for the antiferromagnetic state and the contribution from the AFM order itself, resulting in the cusp-like feature of the AFM state. The direction and magnitude of the spin polarization would be determined by a balance between the incident direction of the X-rays, which favors the *c* direction, and the magnetic anisotropy of $Ni_3TeO_6$, which favors the *ab* direction. This may be the origin of the incident angle dependence of the XCD intensity that is maximized around $\theta \sim 53°$. (See Supporting Information **Figure S8**.)

## 4    Conclusion

In conclusion, we have studied XCD at the Ni $L_{2,3}$ edge of the structurally chiral, collinear antiferromagnet $Ni_3TeO_6$, and observed strong XNCD signals without an external magnetic field. Distinct from XMCD in noncoplanar antiferromagnets, this XNCD is significantly enhanced by antiferromagnetic order and changes sign between domains of opposite crystal chirality. The mechanism of the magnetic enhancement of XNCD is beyond the conventional understanding that most collinear antiferromagnets should exhibit no XMCD because of T symmetry. To explain this new type of XNCD, we propose a model based on symmetry analysis, in analogy to current-induced magnetization in chiral lattice. Our observations provide compelling evidence that supports a novel understanding of the mechanism of XNCD in a chiral material. Our results show that it does not necessarily need to break (effective) time-reversal symmetry to have circular dichroism in magnetic materials. The new mechanism is through the non-vanishing Berry curvature due to the inversion-symmetry broken. This is distinctly different from all previous examples of antiferromagnetic materials that exhibit circular dichroism due to broken time-reversal symmetry. Our results reveal the altermagneism of $Ni_3TeO_6$ and establish the first example of a new class of magnetic materials that can exhibit circular dichroism with time-reversal symmetry. Furthermore, our method can be used as a new spectroscopic tool to study magnetic materials, especially those relevant to chiral spintronics.

# 5 Experimental Section

We conducted Ni $L$-edge soft X-ray spectroscopy and scattering measurements at Beamline 41A of the Taiwan Photon Source.[55] **Figure S1** of the Supporting Information depicts the experimental geometry of our soft-X-ray measurements. The $Ni_3TeO_6$ single crystals used in our study were grown via a flux method; detailed growth conditions are available in Reference.[40] DC magnetic susceptibility measurements, using a magnetic field of $\mu_0H$ = 0.2 T parallel and perpendicular to the $c$-axis, show the typical temperature dependence expected for a collinear antiferromagnet with the spin direction aligned parallel to the $c$-axis. For RXMS, XAS, and circular dichroism measurements, we employed incident X-rays with an energy resolution of 0.3 eV, focused to a lateral size of approximately 1 $\mu$m at the sample position using a capillary tube optic. The scattered X-rays and fluorescence emitted from the $Ni_3TeO_6$ sample to the direction perpendicular to the incident X-rays were detected using a photodiode and a CCD detector. Total fluorescence-yield (TFY) measurements with self-absorption correction[56] were conducted for XAS and circular dichroism. XLD measurements involved with soft X-rays of 50 $\mu$m in vertical and 60 $\mu$m in horizontal, incident on the sample with $\theta$ = 20° off the $a$-axis, were obtained via the partial-fluorescence yield (PFY) method. Additional details can be found in the Supporting Information.

*Statistical Analysis*:
XAS spectral intensities, initially normalized to the incident X-ray flux and after background subtraction, underwent further normalization at 8 eV (5 eV) below the $L_3$ edge and 8 eV (5 eV) above the $L_2$ edge for TFY-XAS (PFY-XAS). RXMS scans were normalized to the average incident X-ray intensity measured by the photodiode before and after the scans. All statistical analyses pertaining to RXMS and XAS intensities were conducted by using Igor Pro software with data errors of ± 1% of the maximum intensity and ± 2% of the $L_3$-edge XAS peak intensity for XCD.

**Supporting Information**
Supporting Information is available from the Wiley Online Library or from the author.

**Acknowledgements**
We acknowledge the contributions of T. A. W. Beale, W. B. Wu, P. D. Hatton, and T. Tyliszczak to the construction of the soft X-ray diffractometer TACoDE. We thank Yusuke Kato and Jun-ichiro Kishine for discussion on CISS effects, and Wei-En Ke for help with magnetization measurements. This work was partly supported by the National Science and Technology Council of Taiwan under Grant No. 1092112-M-213-010-MY3 and 110-2923-M-213-001 and by the Japan Society for the Promotion of Science under Grant No. JP22K03535. The work at Rutgers was supported by W. M. Keck Foundation. A.F. acknowledges the support from the Yushan Fellow Program under the Ministry of Education of Taiwan.

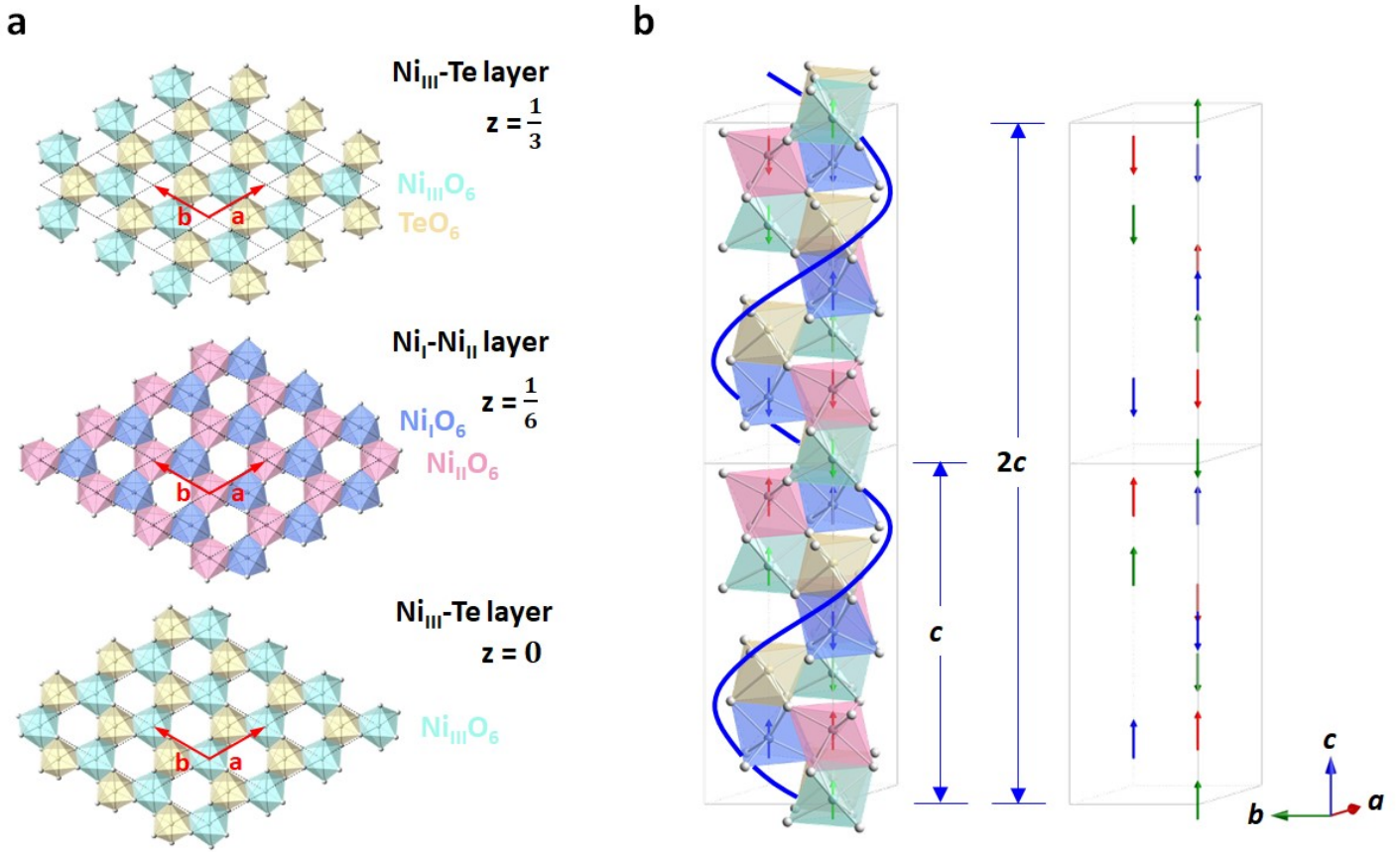

Figure 1: Crystal and spin structures of $Ni_3TeO_6$. a) Two kinds of honeycomb layers in the $ab$ plane of $Ni_3TeO_6$. The $Ni_{III}$-Te layer consists of $Ni_{III}O_6$ and $TeO_6$ octahedra to form a triangular lattice through edge-sharing; the $Ni_I$-$Ni_{II}$ layer comprises $Ni_IO_6$ and $Ni_{II}O_6$ octahedra. $a$ and $b$ are in-plane unit-cell vectors. The alternative stacking of the $Ni_I$-$Ni_{II}$ and $Ni_{III}$-Te layers along the $c$ axis forms the 3D rhombohedral crystal structure. The structural unit cell comprises six honeycomb layers stacked along the $c$ axis. b) Chiral crystal structure of $Ni_3TeO_6$. The blue helix represents the handedness of the crystal chirality. The spins represented by arrows are aligned along the $c$ axis, showing the collinear antiferromagnetic structure of $Ni_3TeO_6$.[33] The magnetic unit cell consists of two structural unit cells, that is, 12 honeycomb layers stacking along the $c$ direction. The structural unit-cell length $c$ and the magnetic unit-cell length $2c$ are indicated by double-headed arrows.

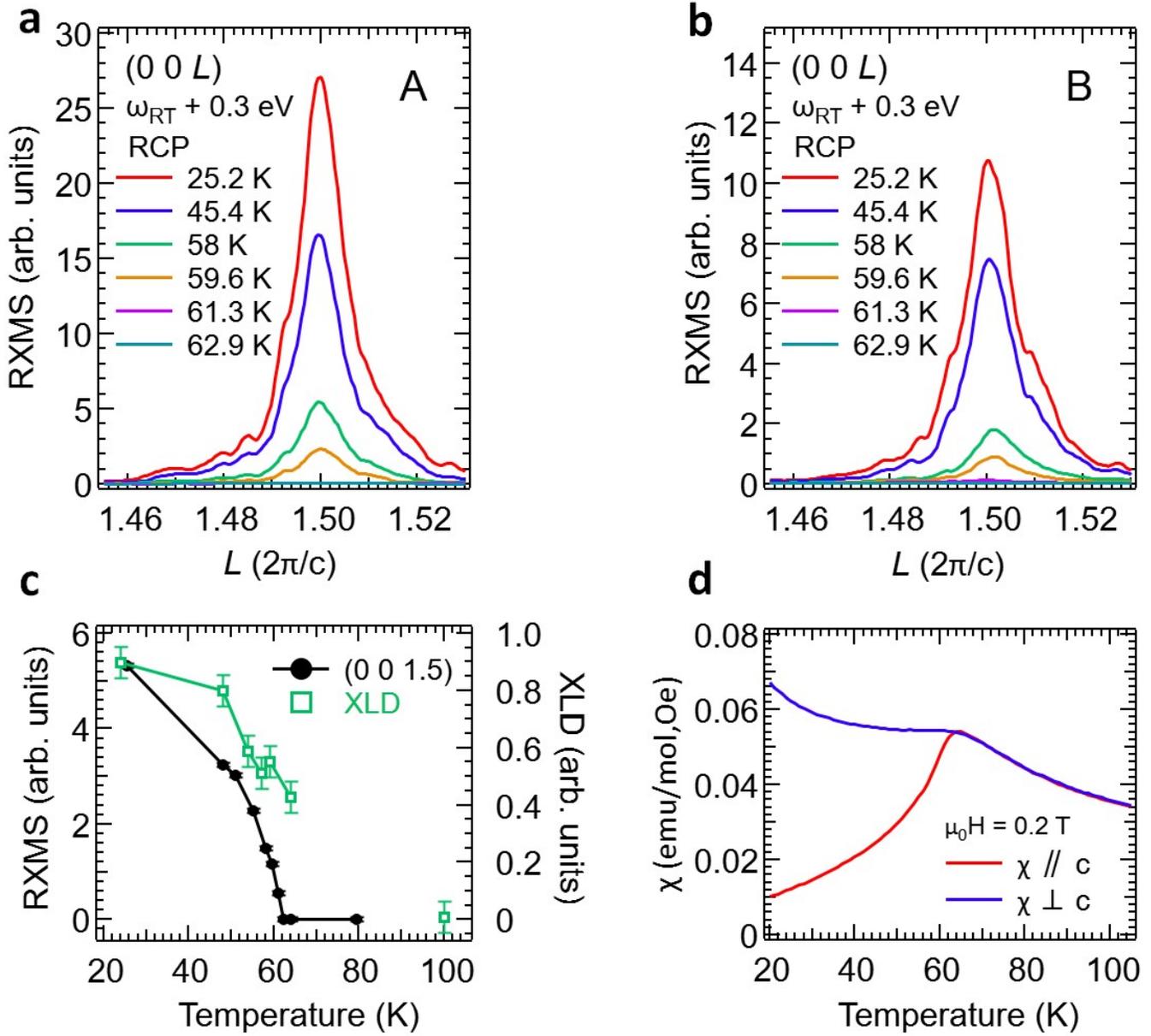

Figure 2: Magnetic transition of $Ni_3TeO_6$. a,b) Resonant X-ray magnetic scattering (RXMS) intensities at $\vec{q}$ = (00$L$) in the reciprocal space scanned along $L$ with RCP at the photon energy of $\omega_{RT}$+0.3 eV at various temperatures at positions A and B, respectively. Here, $\omega_{RT}$ is the Ni $L_3$-edge XAS peak energy defined in Figure 3, and A and B belong to opposite chiral domains shown in Figure 4a. c) Temperature dependence of X-ray linear dichroism (XLD) and the (0, 0, 1.5) peak of RXMS at the Ni $L_3$ edge plotted across the Néel transition. d) DC magnetic susceptibility for magnetic field parallel and perpendicular to the $c$ axis taken with $\mu_0 H$ = 0.2 T.

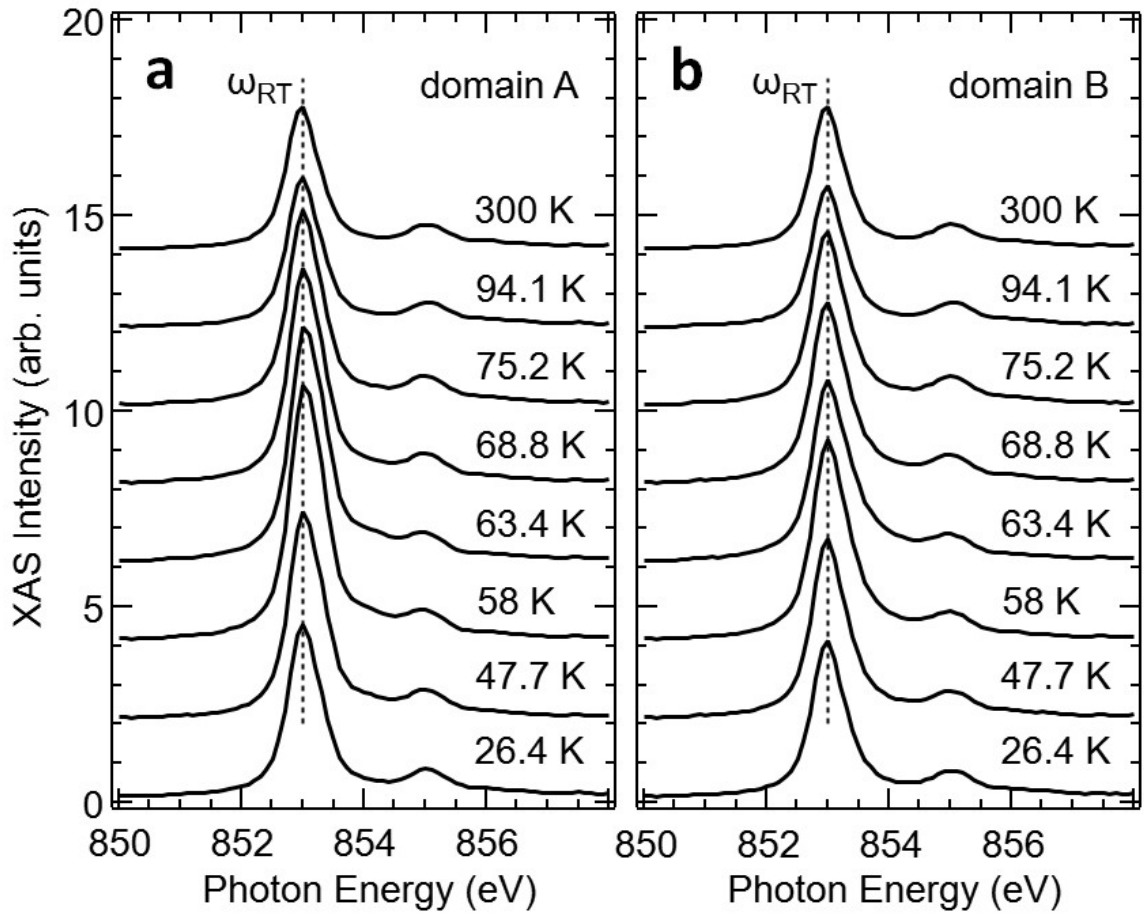

Figure 3: Temperature dependence of Ni $L_3$-edge XAS of $Ni_3TeO_6$. a,b) Ni $L_3$-edge XAS spectra (averaged over RCP and LCP) taken at positions A and B of the sample designated in Figure 4, respectively. The incident angle $\theta$ was ~ 53° off the $a$-axis and fluorescent X-rays from the $Ni_3TeO_6$ sample were detected using a CCD detector. The spectra are vertically shifted for clarity. The vertical broken line shows the photon energy 853 eV ($\equiv \omega_{RT}$) of the $L_3$ peak at room temperature.

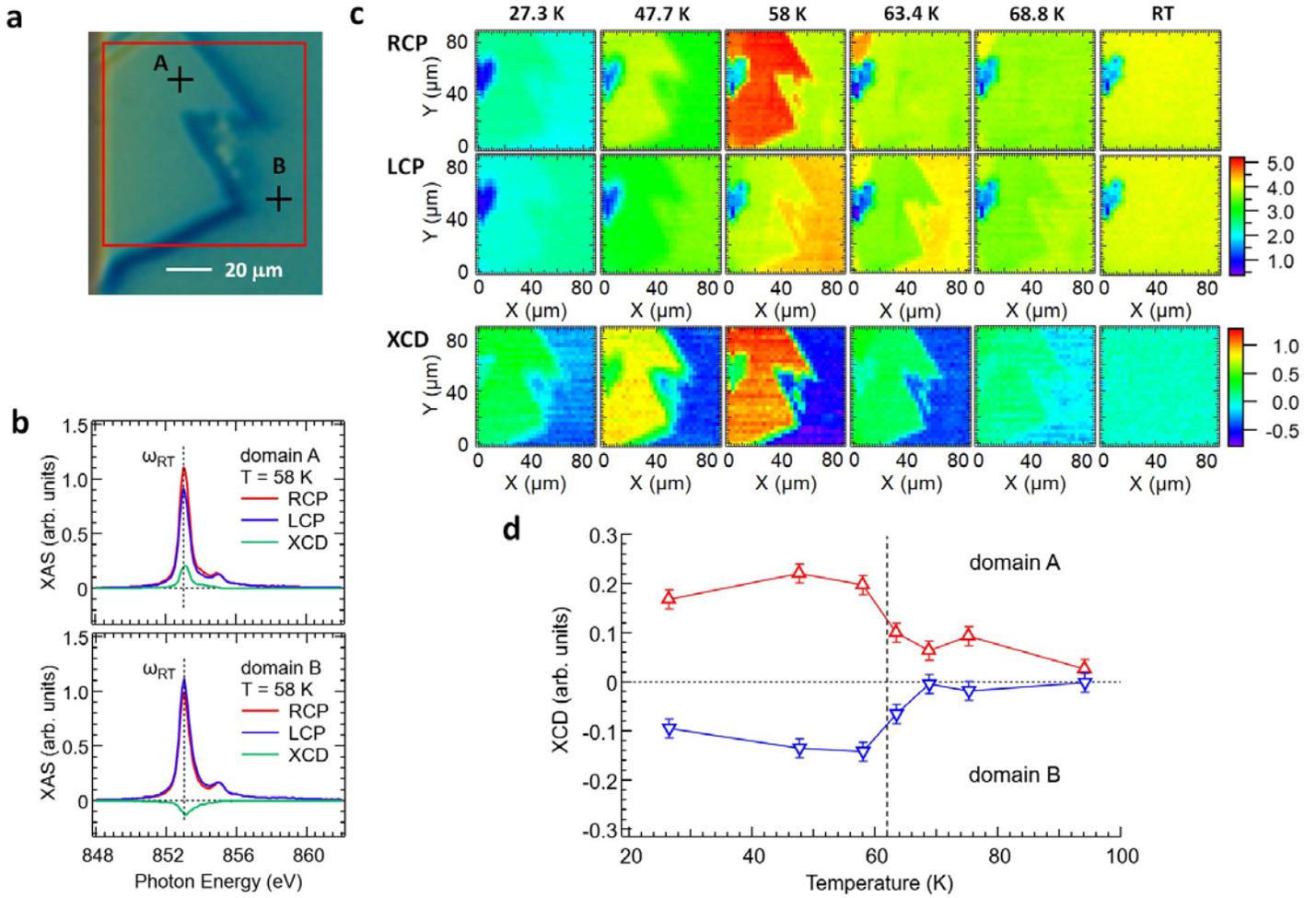

Figure 4: Temperature dependence XNCD of Ni$_3$TeO$_6$. a) Image of the Ni$_3$TeO$_6$ sample recorded with a polarized light microscope. Horizontal and vertical edges are parallel to the [0 1 $\bar{1}$ 0] and [2 $\bar{1}$ $\bar{1}$ 0] axes, respectively. The red square box shows a 90×90-$\mu$m$^2$ area on which the XAS and XCD images were taken. Positions A and B are in two different domains of opposite crystal chirality. b) XAS spectra taken at positions A and B with RCP and LCP and difference spectra, i.e., X-ray circular dichroism (CD) spectra, at 58 K measured by the fluorescence-yield method. Circular dichroism signals at A and B are positive and negative, respectively. c) XAS images of Ni$_3$TeO$_6$ in the 90×90-$\mu$m$^2$ area measured at a photon energy of $\omega_{RT}$+ 0.7 eV (photon energy of $\omega_{RT}$ at RT) with RCP (top) and LCP (middle) and RCP - LCP difference images (bottom) at various temperatures. d) XCD intensity normalized to the $L_3$ XAS peak intensity, defined by 2(RCP LCP)/(RCP + LCP), at positions A and B plotted against temperature. The dichroism changes its sign between A and B.

Supporting Information for

# Giant X-ray circular dichroism in a time-reversal invariant altermagnet

J. Okamoto, R.-P. Wang, Y. Y. Chu, H. W. Shiu, A. Singh, H. Y. Huang,
C. Y. Mou, S. Teh, H. T. Jeng, K. Du, X. Xu, S-W. Cheong, C. H. Du,
C. T. Chen, A. Fujimori, and D. J. Huang

February 9, 2024

This PDF file includes:
- Resonant X-ray magnetic scattering of $Ni_3TeO_6$
- Magnetization curves of $Ni_3TeO_6$
- Non-linear dependence of XCD on the X-ray intensity
- Exchange interactions between Ni ions
- Self-absorption correction on Ni $L_{2,3}$-edge XAS
- X-ray linear dichroism
- Angle-dependent XCD
- X-ray induced orbital magnetization
- XCD and Berry curvature
- Figures S1 to S8

# Resonant X-ray magnetic scattering of $Ni_3TeO_6$

We used a two-circle soft-X-ray diffractometer at the TPS 41A to measure Ni $L$-edge XAS and resonant X-ray magnetic scattering (RXMS) spectra of $Ni_3TeO_6$. Figure S1 depicts the experimental geometry of our soft-X-ray measurements. This diffractometer was constructed by NSRRC in collaboration with Durham university and Tamkang university and named Taiwan Anglo Coherent Diffraction End-station (TACoDE). Incident X-rays were focused to a lateral size of ~1 $\mu$m at the sample position using a capillary tube optic. Scattered X-rays and fluorescence X-rays from the sample were detected by a photodiode and a CCD detector. The sample mounted on a *xyz* stage could be cooled down to 25 K using liquid He. Through piezo motors with feedback signals from laser interferometers, the *xyz* stage controlled the sample position to a precision of 30 nm at low temperatures.

We measured Ni $L_3$-edge RXMS of the collinear antiferromagnetic structure in $Ni_3TeO_6$ with right-handed circularly polarized (RCP) X-rays at scattering vector $\vec{q}$ of (0 0 1.5). Figure S2**b** compares the energy scans of RXMS intensities for domains A (red) and B (blue) at 27.6 K with the Ni $L_3$-edge XAS spectrum at room temperature (RT) (black), which has the maximum absorption intensity at energy $\omega_{RT}$ = 853 eV. XAS spectra of domains A and B overlap at RT as shown in Figure S2**a**. RXMS peaks of A and B were located at $\omega_{RT}$ + 0.3 eV and $\omega_{RT}$ + 0.7 eV, respectively. This difference in line shape reflects the subtle difference in the Ni electronic structures at domains A and B below $T_N$. Next, we measured the temperature dependence of the (0 0 1.5) RXMS intensities at domains A and B. Figures S2**c** and **d** show the (0 0 1.5) RXMS intensities at A and B scanned along $L$ in the reciprocal space at the photon energy of $\omega_{RT}$+0.3 eV at various temperatures. Strong RXMS intensities were observed at both A and B at 25.2 K. As the temperature increased, the RXMS intensities decreased and disappeared above 62 K. Although the (0 0 1.5) RXMS intensities at A was twice as strong as that at B at 25.2 K, the (0 0 1.5) RXMS intensities at A, B, and B' center of the sample surface (the same chiral domain with B) showed similar temperature dependences in Fig. S2**e**. Thus, the Néel temperature $T_N$ of the present $Ni_3TeO_6$ sample is 62 K, which matches with the result of the DC magnetic susceptibility in Fig. 2**d**. Figure S2**f** shows images of the (0 0 1.5) RXMS intensity scanned in the area of Fig. 4**a** at $\omega_{RT}$+0.3 eV with RCP at various temperatures. RXMS intensities after a magnification factor given in the bracket are plotted in a color scale at each temperature. Below 58 K, domains are clearly distinguished by different colors. The intensity of domain A is stronger than that of domain B below 58 K, but becomes similar at 59.6 K. It reverses at 61.3 K and no clear domain can be seen at 62.9 K. These results also indicate that $T_N$ of the present $Ni_3TeO_6$ is 62 K. The weak-intensity area in domain A near (X, Y) = (5, 70) is an area of degraded crystal quality.

## Magnetization curves of $Ni_3TeO_6$

Figures S3**a** and S3**b** show the magnetization curves of $Ni_3TeO_6$ for H ∥ *c* and H ⊥ *c* at various temperatures, respectively. Neither hysteresis nor remanence was observed for both field directions. For H ∥ *c* above the critical field $H_c \sim$ 6-9 T, the Ni spins flip from the *c* direction to the *ab* plane, becoming incommensurate and spiral with canting in the *c* direction as shown in Fig. S3**a** [40].

## Non-linear dependence of XCD on the X-ray intensity

Figure S4 shows the XAS spectra with RCP and LCP and XCD spectra of $Ni_3TeO_6$ taken at positions A and B in Fig. 4**a** at 26 K. XAS and XCD intensities are normalized to the $L_3$ average XAS peak intensity, (RCP+LCP)/2. As the incident X-ray intensity was decreased to 60%, it was observed that normalized XCD intensity decreased to $\sim \frac{1}{2}$, as shown by the dotted curve XCD'.

## Exchange interactions between Ni ions

Figure S5 defines the spin exchange constants $J_1 - J_5$ between $Ni^{2+}$ ($S$ = 1) ions in $Ni_3TeO_6$. Wu et al. calculated them based on spin-polarized density functionals using the GGA+$U$ method and found that $J_1 = -0.94$ meV and $J_2 = -4.55$ meV for $U$ = 2.5 eV [44]. The exchange constant $J_2$ is strongly ferromagnetic (FM), but $J_1$ is weakly FM. In addition, $J_3$ = 3.70 meV, $J_4$ = 6.41 meV, and $J_5$ = 1.48 meV for $U$ = 2.5 eV. That is, the exchange constants $J_3$ and $J_4$ are strongly antiferromagnetic (AFM) but $J_5$ is weakly AFM. These exchange constants explain the observed magnetic structure below $T_N$, as illustrated in Fig. 1**b**.

## Self-absorption correction on Ni $L_{2,3}$-edge XAS

We investigated the influence of the self-absorption effect on the Ni $L_{2,3}$-edge XAS spectra of $Ni_3TeO_6$ measured in the fluorescence-yield method (FY-XAS). Figure S6**a** shows the measured FY-XAS spectra (averaged over RCP and LCP and positions A and B) at 63 K for the incident angles of $\theta$ = 20, 53, and 75°. The broken blue curve is FY-XAS measured at $\theta$ = 53° and RT. We calculated the XAS spectra with the self-absorption effect corrected referring to Eisebitt's method [56]: Using two FY-XAS spectra with different experimental geometries, one can correct for the self-absorption effect in the FY-XAS spectra. Figure S6**b** compares the self-absorption-corrected XAS spectra: The red curve is obtained from FY-XAS at $\theta$ = 20 and 75° shown in Fig. S6**a** and the blue one is obtained from the FY-XAS spectrum of $\theta$ = 53° by multiplying a correction factor estimated from XAS spectra of $\theta$ = 53° at RT. There exists a clear difference between these corrected XAS spectra, which reveals that the enhancement of XAS (and X-ray CD) at $\theta \sim$ 53° below $T_N$ is independent of the self-absorption effect on FY-XAS spectra. Figure S6**c** shows self-absorption corrected Ni $L_{2,3}$-edge XAS spectra at positions A and B with RCP, LCP, and their difference spectra (XCD) of $\theta$ = 53° at 58 K. Large XCD intensities are observed at $L_3$ edge as in Figure 4**b** but those at $L_2$ edge are negligibly small.

## X-ray linear dichroism

Figure S7**a** shows XAS spectra of incident X-rays with electric field vector parallel ($I_\parallel$) and perpendicular to the $c$ axis ($I_\perp$) and their difference, i.e., X-ray linear dichroism (XLD) at 24 K. $I_\parallel$ and $I_\perp$ spectra were obtained by XAS spectra measured with $\pi$ ($I_\pi$) and $\sigma$ ($I_\sigma$)

polarizations in Fig. S1: $I_\perp = I_\sigma$ and $I_\parallel = \frac{1}{\cos^2\theta}(I_\pi - \sin^2\theta I_\perp)$ at incident angle $\theta$ of 20°. The vanishingly small but finite XLD at temperatures well above $T_N$, which is less than 2% of $L_3$-edge XAS peak intensity at 100 K, originates from the $Ni^{2+}$ ions in a crystal field of $C_{3v}$ symmetry. The enhanced XLD below $T_N$, which is ~ 20% of $L_3$-edge XAS peak intensity at 24 K, is dominated by the contributions from the collinear AFM ordering. By incorporating $p$-to-$d$ charge-transfer effects, our multiplet calculations can reproduce the spectral lineshapes of XAS and XLD, as illustrated in Fig. S7**b**. The temperature dependence of XLD intensity is consistent with that of RXMS, as plotted in Fig. 2**c**.

## Angle-dependent XCD

We investigated the angle dependence of X-ray circular dichroism (XCD) by measuring Ni $L_3$-edge XAS images with the incident photon energy of $\omega_{RT}$+0.3 eV across $T_N$. Figure S8**a** reveals that XAS images at the incident angle $\theta$ = 53° show a clear contrast between different chiral domains, but those at $\theta$ = 20 and 75° show a very weak contrast at 62 and 46 K. No clear contrast was observed in the XAS images at $\theta$ = 90°. Ni $L_3$-edge XAS and difference spectra at various incident angles plotted in Figs. S8**b** and S8**c** also show very small XCD intensities at 62 K for $\theta$ = 20 and 75°, and no XCD intensities observed at $\theta$ = 90°. These results indicate that the emergent XCD at the Ni $L_3$-edge of $Ni_3TeO_6$ shows a strong incident angle dependence, in agreement with the scenario of spin-chirality coupling and anisotropic magnetic susceptibility of $Ni_3TeO_6$ discussed in the main text.

## X-ray induced orbital magnetization

In the presence of X-ray, to the linear term, the Hamiltonian can be written as

$$H = H_0 - \int d^3r\, \vec{j}\cdot\vec{A}, \tag{S1}$$

where $H_0$ is the unperturbed Hamiltonian, $\vec{A}$ is the vector potential of the incident X-ray, and $\vec{j}$ is the electric current operator. By using the Helmholtz's theorem, $\vec{j}$ can be generally decomposed as $\vec{J} = -\nabla \phi_J + \nabla \times \vec{M}$. Here because $\nabla \cdot \vec{J} = -\nabla^2 \phi_J$ and $\nabla \cdot \vec{J} = -\frac{\partial}{\partial t}\rho$ (continuity equation), it is clear that $\phi_J$ is determined by the charge density $\rho$. On the other hand, $\vec{M}$ is given by

$$\vec{M} = \frac{1}{4\pi} \int \frac{\nabla' \times \vec{J}(\vec{r}')}{|\vec{r} - \vec{r}'|} d^3\vec{r}'. \tag{S2}$$

By using the identity $\nabla' \times \vec{J}(\vec{r}')/|\vec{r} - \vec{r}'| = \nabla' \times (\vec{J}(\vec{r}')/|\vec{r} - \vec{r}'|) - \nabla'(1/|\vec{r} - \vec{r}'|) \times \vec{J}(\vec{r}')$, one finds

$$\vec{M}(\vec{r}) = \frac{1}{4\pi} \int \frac{(\vec{r}' - \vec{r}) \times \vec{J}(\vec{r}')}{|\vec{r} - \vec{r}'|^3} d^3\vec{r}', \tag{S3}$$

where the integration over the term $\nabla' \times (\vec{J}(\vec{r}')/|\vec{r} - \vec{r}'|)$ is converted to a surface integral at infinity and can be neglected. Now considering the contribution of $\vec{J}$ to $\vec{M}$ locally, we set $\vec{r}'$ close to $\vec{r}$ by setting $\vec{r}' = \alpha \vec{r}$ with $\alpha \to 1^+$ so that $\vec{J}(\vec{r}')$ in Eq. (S3) can be taken outside the integration with $\vec{r}'$ being replaced by $\vec{r}$. Hence we obtain

$$\vec{M}(\vec{r}) = \left( \frac{1}{4\pi} \int \frac{1}{|\vec{r} - \vec{r}'|^2} d^3\vec{r}' \right) \hat{r} \times \vec{J}(\vec{r}), \tag{S4}$$

where the integration of $\vec{r}'$ is over the forward direction with the polar angle in spherical coordinates being in the region $0 \leq \theta \leq \pi/2$. After the integration of $\vec{r}'$, we obtain

$$\vec{M} = \frac{1}{2} \vec{r} \times \vec{J}. \tag{S5}$$

In other words, the operator $\vec{M}$ is the orbital magnetization and $\nabla \times \vec{M}$ corresponds to the magnetization current.

The contribution of the orbital magnetization in Eq.(S1) can thus be written as

$$H_M = -\int d^3\vec{r}\, \vec{A} \cdot \nabla \times \vec{M}. \tag{S6}$$

Finally, after integration by parts, we obtain

$$H_M = -\int d^3\vec{r}\, \nabla \times \vec{A} \cdot \vec{M} = -\int d^3\vec{r}\, \vec{B} \cdot \vec{M}. \tag{S7}$$

The above equation implies that the X-ray will induce orbital magnetization given by the average of $\vec{M}$, $\langle \vec{M} \rangle$. When the incident direction $\vec{k}$ of the x-ray is reversed to $-\vec{k}$, the Poynting vector $\vec{E} \times \vec{B}$ is also reversed. This can be viewed as $\vec{B}$ goes to $-\vec{B}$. Hence the induced orbital magnetization $\langle \vec{M} \rangle$ also changes the sign. For $\vec{k}$ being parallel to the axial direction of the chiral structure that only breaks the mirror symmetry, the orbital current in the chiral structure is reversed with the same magnitude. Hence the reversed orbital magnetization has magnitude so that $\vec{M}$ simply goes to $-\vec{M}$. The configurations of $\vec{B}$ and $\vec{M}$ for $\vec{k}$ is the same as that for $-\vec{k}$.

Hence the observed XNCD is reciprocal.

## XCD and Berry curvature

**Berry curvature**

Consider a quantum mechanical system in which the Hamiltonian depends on time through a set of time-dependent parameters denoted by $\vec{R}(t) = (R_1, R_2, ..., R_N)$, i.e.,

$$\hat{H} = \hat{H}(\vec{R}). \tag{S8}$$

If $|n(\vec{R})\rangle$ form an orthonormal basis of $\hat{H}(\vec{R})$,

$$\hat{H}|n(\vec{R})\rangle = E_n(\vec{R})|n(\vec{R})\rangle. \tag{S9}$$

According to the quantum adiabatic theorem, if $|\psi_n(t)\rangle$ is an eigenstate of $\hat{H}(t)$, the general solution to the time-dependent Schrodinger equation

$$i\hbar \frac{\partial}{\partial t}|\psi_n(t)\rangle = \hat{H}|\psi_n(t)\rangle \tag{S10}$$

includes two phase factors, i.e.,

$$|\psi_n(t)\rangle = e^{i\theta_n(t)} e^{i\gamma_n(t)} |\psi_n(t)\rangle, \tag{S11}$$

where $\theta_n(t)$ is referred to as the dynamical phase,

$$\theta_n(t) = -\frac{1}{\hbar} \int_0^t E_n(t') dt'. \tag{S12}$$

And $\gamma_n(t)$ is the geometric phase. One can prove that

$$\gamma_n(t) = i \int_0^t \langle \psi_n(t') | \dot{\psi}_n(t') \rangle \, dt' \tag{S13}$$

$$= i \int_{R_i}^{R_f} \langle \psi_n | \nabla_R \psi_n \rangle \cdot d\vec{R}, \tag{S14}$$

where $\nabla_R$ is the gradient with respect to each time-dependent parameter $\vec{R}$. If the Hamiltonian returns to its original form through a closed path in the parameter space, the total geometric phase

$$\gamma_n(t) = \oint \mathcal{A}_n \cdot d\vec{R} \tag{S15}$$

is known as the Berry phase, and $\mathcal{A}_n \equiv i \langle \psi_n | \nabla_R \psi_n \rangle$ is called the Berry connection or the Berry vector potential. Using Stokes' theorem, we have

$$\gamma_n(t) = \int (\nabla_R \times \mathcal{A}_n) \cdot d\vec{S}, \tag{S16}$$

and $\vec{\Omega_n} \equiv \nabla_R \times \mathcal{A}_n$ is called the Berry curvature. In analogy to electromagnetism, Berry phase, Berry connection, and Berry curvature correspond to magnetic flux, vector potential and magnetic field, respectively. If the parameter space is two dimensional, we have

$$\vec{\Omega}(\vec{k}) = i \langle \boldsymbol{\nabla}_k \psi_n | \times | \boldsymbol{\nabla}_k \psi_n \rangle \tag{S17}$$

and

$$\Omega_z^{\vec{k}} = i \left( \langle \frac{\partial \psi_n}{\partial k_x} | \frac{\partial \psi_n}{\partial k_y} \rangle - \langle \frac{\partial \psi_n}{\partial k_y} | \frac{\partial \psi_n}{\partial k_x} \rangle \right) \tag{S18}$$

**Symmetry properties of $\vec{\Omega}(\vec{k})$**

If the system is invariant under space inversion I,

$$I\vec{\Omega}(\vec{k}) = \vec{\Omega}(-\vec{k}) = \vec{\Omega}(\vec{k}), \tag{S19}$$

and therefore $\vec{\Omega}(\vec{k})$ is an even function of $\vec{k}$. Under under time-reversal transformation T,

$$T\vec{\Omega}(\vec{k}) = -\vec{\Omega}(-\vec{k}), \tag{S20}$$

because T is an anti-unitary transformation. That is, if the system has time-reversal symmetry,

$$\vec{\Omega}(\vec{k}) = -\vec{\Omega}(-\vec{k}) \tag{S21}$$

That is, $\vec{\Omega}(\vec{k})$ is an odd function of $\vec{k}$, and the integration of $\vec{\Omega}(\vec{k})$ in the momentum space vanishes. Hence, if you have time reversal T and space inversion I symmetry, the Berry curvature vanishes identically at all k-points.

**Circular dichroism in X-ray absorption**

The transition rate of X-ray absorption (XAS) from an initial state $|\psi_i\rangle$ of energy $E_i$ to a final state $|\psi_f\rangle$ of energy $E_f$ is

$$\Gamma_{\text{XAS}}(\hbar\omega) = \frac{2\pi}{\hbar} \sum_f \left| \langle \psi_f | \frac{e}{mc} \vec{A} \cdot \vec{P} | \psi_i \rangle \right|^2 \delta(E_f - E_i - \hbar\omega), \tag{S22}$$

where $\hbar\omega$ and $\vec{\epsilon}$ are, respectively, the energy and the polarization of the incident X-ray, and the vector potential $\vec{A}$ is

$$\vec{A} = A_0 e^{i(\vec{k}\cdot\vec{r}-\omega t)} \vec{\epsilon}, \quad A_0 = \sqrt{\frac{2\pi c^2 \hbar N}{\omega V}}. \tag{S23}$$

Using the relation $[H, \vec{r}] = \frac{\hbar}{i} \frac{\vec{P}}{m}$ and the dipole approximation $e^{i\vec{k}\cdot\vec{r}} \approx 1$, it is straightforward to show that

$$\langle \psi_f | e^{i\vec{k}\cdot\vec{r}} \vec{\epsilon} \cdot \vec{P} | \psi_i \rangle \approx im \frac{E_f - E_i}{\hbar} \langle \psi_f | \vec{\epsilon} \cdot \vec{r} | \psi_i \rangle = im\omega \langle \psi_f | \vec{\epsilon} \cdot \vec{r} | \psi_i \rangle. \tag{S24}$$

The Poynting vector is $\vec{S} = \frac{c}{4\pi} \vec{E} \times \vec{B}$ and $S = \frac{\omega^2}{4\pi c} A_0^2$. The photon flux is $S/\omega = \frac{\omega}{4\pi c} A_0^2$. Therefore the XAS cross section in the dipole approximation is

$$\sigma_{\text{XAS}}(\hbar\omega) = \frac{4\pi\hbar c}{\omega A_0^2} \Gamma_{\text{XAS}} \approx \frac{4\pi c}{A_0^2} \frac{2\pi e^2}{\hbar c^2} A_0^2 \sum_{i,f} (\hbar\omega) \left| \langle \psi_f | \vec{\epsilon} \cdot \vec{r} | \psi_i \rangle \right|^2 \delta(E_f - E_i - \hbar\omega), \tag{S25}$$

If $\vec{\epsilon}_\pm = \frac{\hat{\epsilon}_x \pm i\hat{\epsilon}_y}{\sqrt{2}}$,

$$\vec{\epsilon} \cdot \hat{r} = \vec{\epsilon}_- \cdot \left(\frac{\hat{x}+i\hat{y}}{\sqrt{2}}\right) + \vec{\epsilon}_+ \cdot \left(\frac{\hat{x}-i\hat{y}}{\sqrt{2}}\right).$$

For incident X-rays along $\hat{\vec{z}}$ with circular polarizations $\vec{\epsilon}_\pm$, the spectrum of X-ray circular dichroism (XCD) in XAS of the 2p→3d transition is

$$\sigma_{\text{XCD}}(\hbar\omega) \approx \frac{2\pi^2 e^2}{\hbar c}\hbar\omega \sum_f \left( \left|\langle\psi_f|(\hat{\epsilon}_x - i\hat{\epsilon}_y)\cdot(\hat{x}+i\hat{y})|\psi_i\rangle\right|^2 - \left|\langle\psi_f|(\hat{\epsilon}_x + i\hat{\epsilon}_y)\cdot(\hat{x}-i\hat{y})|\psi_i\rangle\right|^2 \right) \delta(E_f - E_i - \hbar\omega)$$

(S26)

Using the relations $\langle\psi_f|\hat{x}|\psi_i\rangle = \left\langle -i\frac{\partial}{\partial k_x}\psi_f \middle| \psi_i \right\rangle$ and $\langle\psi_f|\hat{y}|\psi_i\rangle = \left\langle -i\frac{\partial}{\partial k_y}\psi_f \middle| \psi_i \right\rangle$ we can show that

$$\left|\langle\psi_f|\hat{x}\pm i\hat{y}|\psi_i\rangle\right|^2 = \left|\langle\tfrac{\partial}{\partial k_x}\psi_f|\psi_i\rangle\right|^2 + \left|\langle\tfrac{\partial}{\partial k_y}\psi_f|\psi_i\rangle\right|^2 \mp i\left(\langle\tfrac{\partial}{\partial k_x}\psi_f|\psi_i\rangle\langle\psi_i|\tfrac{\partial}{\partial k_y}\psi_f\rangle - \langle\tfrac{\partial}{\partial k_y}\psi_f|\psi_i\rangle\langle\psi_i|\tfrac{\partial}{\partial k_x}\psi_f\rangle\right),$$

and then have

$$\sigma_{\text{XCD}}(\hbar\omega) \approx -\frac{4\pi^2 e^2}{\hbar c}\hbar\omega \sum_f \left( \langle\tfrac{\partial}{\partial k_x}\psi_f|\psi_i\rangle\langle\psi_i|\tfrac{\partial}{\partial k_y}\psi_f\rangle - \langle\tfrac{\partial}{\partial k_y}\psi_f|\psi_i\rangle\langle\psi_i|\tfrac{\partial}{\partial k_x}\psi_f\rangle \right) \delta(E_f - E_i - \hbar\omega).$$

(S27)

**Approximation: neglecting the core-hole potential**

For a core hole of lifetime width Γ which is much smaller than the X-ray energy, $\hbar\omega \gg \Gamma$, we approximate the Dirac delta function by using the Lorentzian function,

$$\delta(E_f - E_i - \hbar\omega) \approx \frac{\Gamma/2\pi}{(E_f - E_i - \hbar\omega)^2 + \Gamma^2/4},$$

(S28)

Assuming that the initial state $|\psi_i\rangle$ is composed of the $N$-electron 3$d$ states $|\phi_N^{k_i}\rangle$ of wave vector $k_i$ and the 2$p$ core level of the $m$-th atom is completely filled, and the final state $|\psi_f\rangle$ is the ($N$ + 1)-electron 3$d$ states accompanied by a 2$p$ core-hole state $|\phi_c^m\rangle$, i.e.,

$$|\psi_i\rangle = |\phi_{n_i,k_i}^N\rangle$$

(S29)

$$|\psi_f\rangle = |\phi_{n,k}^{N+1}\phi_c^m\rangle ,$$

(S30)

where $n$ is the band index. Also, $E_i + \hbar\omega \equiv E + \hbar\omega_0$ and $E_f = E_{n,k} + \hbar\omega_0$, where $E > 0$, $\hbar\omega_0$ is the 2$p$ binding energy, and $E_k$ is the energy of the single-electron state of wave vector $k$.

Then the XCD spectrum is

$$\sigma_{\text{XCD}}(\hbar\omega_0 + E) \approx -\frac{4\pi^2 e^2}{\hbar c}\hbar\omega \sum_{n_i,k_i,n,k} \frac{\langle\tfrac{\partial}{\partial k_x}\phi_k^{N+1}\phi_c^m|\phi_N^{k_i}\rangle\langle\phi_N^{k_i}|\tfrac{\partial}{\partial k_y}\phi_{n,k}^{N+1}\phi_c^m\rangle - \langle\tfrac{\partial}{\partial k_y}\phi_{n,k}^{N+1}\phi_c^m|\phi_N^{k_i}\rangle\langle\phi_N^{k_i}|\tfrac{\partial}{\partial k_x}\phi_{n,k}^{N+1}\phi_c^m\rangle}{(E_{n,k}-E)^2+\Gamma^2/4}.$$

(S31)

As the 2$p$ binding energy $\hbar\omega_0$ is independent of site, we apply the completeness relation

$$\sum_{n_i,k_i} |\phi_{n_i,k_i}^N\rangle \langle \phi_{n_i,k_i}^N| = 1 \tag{S32}$$

and get

$$\sigma_{\text{XCD}}(\hbar\omega_0 + E) \approx -\frac{4\pi^2 e^2}{\hbar c}\hbar\omega \sum_{n,k} \frac{\langle \frac{\partial}{\partial k_x}\phi_{n,k}^{N+1}\phi_c^m|\frac{\partial}{\partial k_y}\phi_{n,k}^{N+1}\phi_c^m\rangle - \langle \frac{\partial}{\partial k_y}\phi_{n,k}^{N+1}\phi_c^m|\frac{\partial}{\partial k_x}\phi_{n,k}^{N+1}\phi_c^m\rangle}{(E_{n,k} - E)^2 + \Gamma^2/4}.$$

(S33)

The presence of the core-hole state $\phi_c^m\rangle$ yields an effect similar to that of a core-hole potential. Assume that the core-hole potential does not change the Berry curvature. Neglecting this potential, we have the following the approximations:

$$\langle \frac{\partial}{\partial k_x}\phi_{n,k}^{N+1}\phi_c^m|\frac{\partial}{\partial k_y}\phi_{n,k}^{N+1}\phi_c^m\rangle \approx \langle \frac{\partial}{\partial k_x}\phi_{n,k}^{N+1}|\frac{\partial}{\partial k_y}\phi_{n,k}^{N+1}\rangle \langle \phi_c^m|\phi_c^m\rangle \tag{S34}$$

$$\langle \frac{\partial}{\partial k_y}\phi_{n,k}^{N+1}\phi_c^m|\frac{\partial}{\partial k_x}\phi_{n,k}^{N+1}\phi_c^m\rangle \approx \langle \frac{\partial}{\partial k_y}\phi_{n,k}^{N+1}|\frac{\partial}{\partial k_x}\phi_{n,k}^{N+1}\rangle \langle \phi_c^m|\phi_c^m\rangle . \tag{S35}$$

Dropping the index $N + 1$, we have an expression of XCD in terms of the Berry curvature:

$$\sigma_{\text{XCD}}(\hbar\omega_0 + E) \approx -\frac{4\pi^2 e^2}{\hbar c}\hbar\omega_0 \sum_{n,k} \frac{\langle \frac{\partial}{\partial k_x}\phi(k)|\frac{\partial}{\partial k_y}\phi(k)\rangle - \langle \frac{\partial}{\partial k_y}\phi(k)|\frac{\partial}{\partial k_x}\phi(k)\rangle}{(E_{n,k} - E)^2 + \Gamma^2/4} \tag{S36}$$

Therefore,

$$\sigma_{\text{XCD}}(\hbar\omega_0 + E) \approx -\frac{4\pi^2 e^2}{\hbar c}\hbar\omega_0 \sum_n \int_{BZ} \frac{d^3\vec{k}}{(2\pi)^3} \frac{\Omega_z^{\vec{k}}}{(E_{n,k} - E)^2 + \Gamma^2/4}. \tag{S37}$$

**Sum rule**

We integrate the XCD over the $L_{2,3}$ absorption edges and have

$$\int \sigma_{\text{MCD}}(\hbar\omega)d\hbar\omega = -\frac{8\pi^2 e^2}{\hbar c} \int_{L_{2,3}} d\hbar\omega \int_{BZ} \hbar\omega\delta(E_k + \hbar\omega_0 - \hbar\omega)\Omega_z^{\vec{k}} \frac{d^3\vec{k}}{(2\pi)^3}$$

$$= -\frac{8\pi^2 e^2}{\hbar c} \int_{BZ} (E_k + \hbar\omega_0)\Omega_z^{\vec{k}} \frac{d^3\vec{k}}{(2\pi)^3}$$

$$\approx -\frac{8\pi^2 e^2}{\hbar c}\omega_0 \int_{BZ} \Omega_z^{\vec{k}} \frac{d^3\vec{k}}{(2\pi)^3},$$

(S38)

because $\hbar\omega_0 \gg E_k$. Combining the XCD sum rule of orbital magnetic moment per ion [11]

$$m_{\text{orb}} = -\frac{4 \int_{L_{2,3}} \sigma_{\text{MCD}}(\hbar\omega)d\omega}{3 \int_{L_{2,3}} \sigma_{\text{XAS}}(\hbar\omega)d\omega}(10 - n_d), \tag{S39}$$

where $n_d$ is the number of 3d electrons per Ni ion, we have

$$m_{\text{orb}} \propto \frac{\int_{BZ} \Omega_z(k) d^3\vec{k}}{\int_{L_{2,3}} \sigma_{\text{XAS}}(\hbar\omega)d\omega}. \tag{S40}$$

This is consistent with the orbital magnetization arising from the Berry phase correction to the electron density of states [47].

$$M_{\text{orb}} = \frac{e}{\hbar} \int^{\mu_0} \frac{d^3\vec{k}}{(2\pi)^3} \vec{\Omega}[\mu_0 - \epsilon(k)] \tag{S41}$$

We now have a consistent picture. This new expression of XCD in terms of Berry curvature shows that XCD can unravel the orbital magnetization hidden in an antiferromagnetic spin structure in which the integration of Berry curvature does not vanish, although the system does not have a net spin moment. In the present study, it has been difficult to extend the normalization of the XAS spectra to the $L_2$-edge region. Therefore, the orbital sum rule (S41) remains to be tested in future studies.

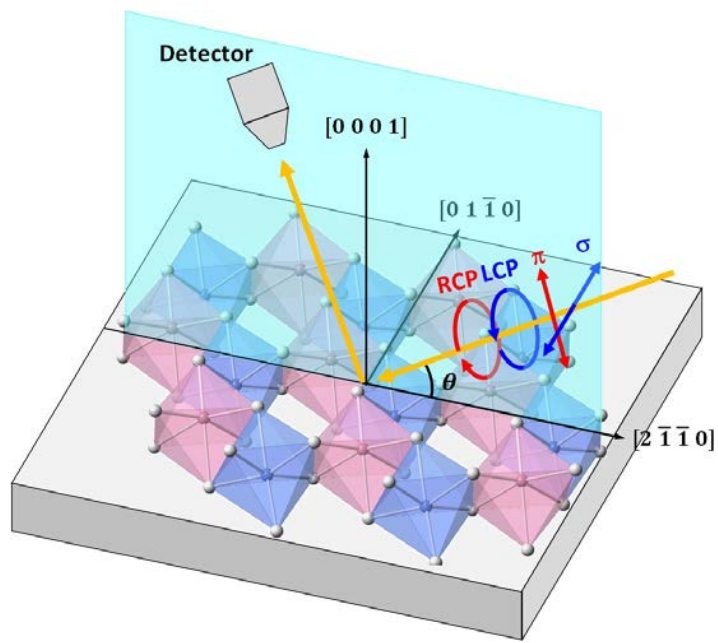

Figure S1: **Illustration of the measurement geometry for XAS, RXMS, and RIXS.** The sample surface was parallel to the *ab* plane. The scattering plane was set to the $[2, \bar{1}, \bar{1}, 0]$ – $[0,0,0,1]$ plane. Soft X-rays with $\pi$, $\sigma$ ($\epsilon \perp c$), RCP, or LCP polarization were incident on the Ni$_3$TeO$_6$ sample with an angle $\theta$ off the $[2, \bar{1}, \bar{1}, 0]$ direction, i.e, the hexagonal *a* axis. Fluorescent X-rays and scattered X-rays from Ni$_3$TeO$_6$ were detected using a photodiode, a CCD, or the AGMAGS spectrometer for RIXS [55].

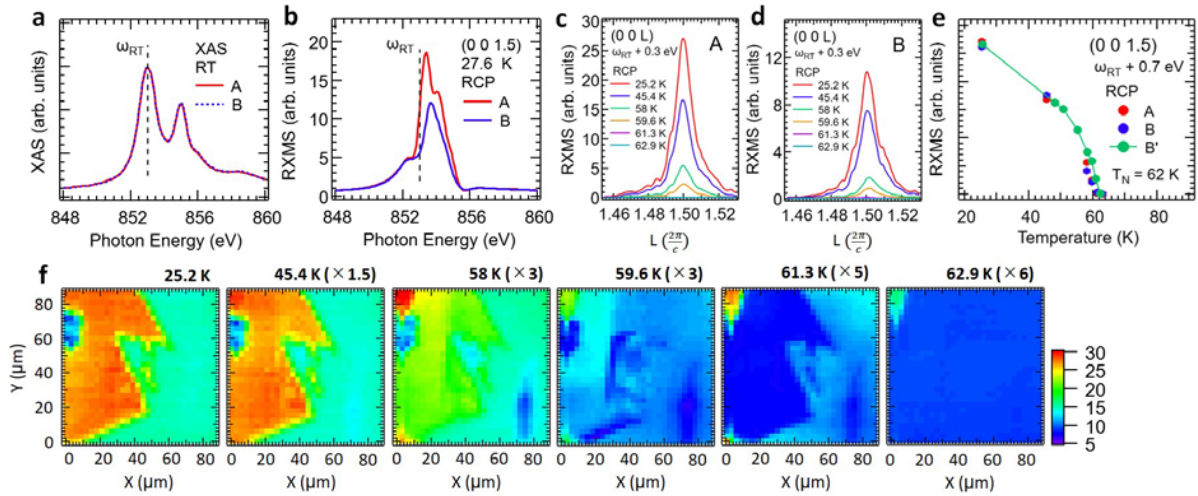

Figure S2: **Ni $L_3$-edge resonant X-ray magnetic scattering of $Ni_3TeO_6$ with $\vec{q}$ = (0 0 1.5).** **a**, Ni $L_3$-edge XAS spectra of domain A and B at RT. **b**, RXMS spectra measured with RCP at 27.6 K. Red and blue curves are energy scans of domains A and B, respectively. The vertical dashed line shows the X-ray energy $\omega_{RT}$ (853 eV) of the XAS peak at RT. The RXMS peaks of domain A and B are at $\omega_{RT}$+0.3 eV and $\omega_{RT}$+0.7 eV, respectively. **c, d**, RXMS intensities with RCP scanned along $L$ in the reciprocal space at the photon energy $\omega_{RT}$+0.3 eV at selected temperatures at A and B, respectively. **e**, Temperature-dependent RXMS intensity at A, B, and B', which is another position of chiral domain B. The photon energies were set to $\omega_{RT}$+0.3 eV for A and B and $\omega_{RT}$+0.7 eV for B'. **f**, Images of RXMS intensity scanned in the area of Fig. 4a at $\omega_{RT}$+0.3 eV with RCP at selected temperatures. RXMS intensities after a magnification factor given in the bracket are plotted in a color scale at each temperature. The weak-intensity area in domain A near (X, Y) = (5, 70) is an area of degraded crystal quality.

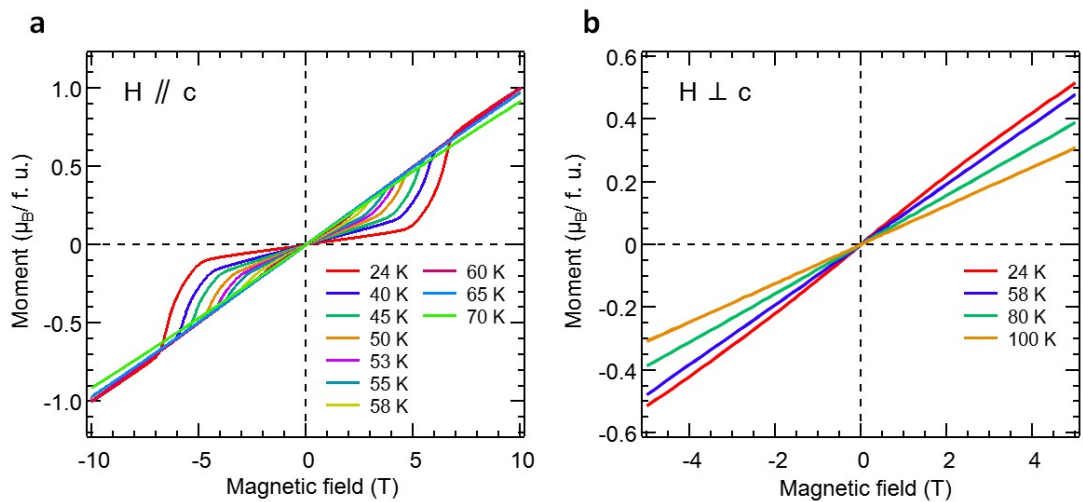

Figure S3: **Magnetization curves of $Ni_3TeO_6$ across $T_N$. a, b** Magnetization curves for the magnetic field applied parallel (**a**) and perpendicular (**b**) to the $c$ axis at various temperatures, respectively.

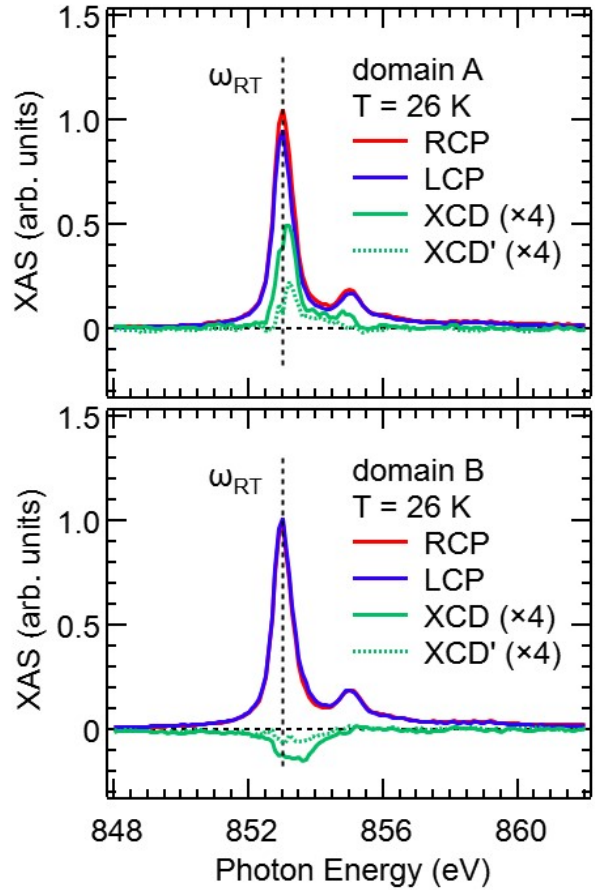

Figure S4: **Non-linear dependence of XCD of $Ni_3TeO_6$ on the X-ray intensity.** XAS spectra with RCP and LCP and XCD spectra taken at positions A and B at 26 K. XAS and XCD intensities are normalized to the $L_3$ average XAS peak intensity, (RCP+LCP)/2. As the incident X-ray intensity was decreased to 60 %, the normalized XCD intensity (XCD') decreased to $\sim\frac{1}{2}$.

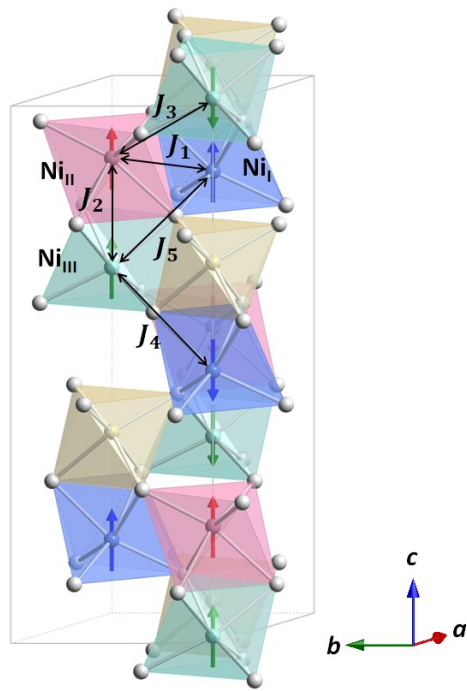

Figure S5: **Exchange interactions between $Ni^{2+}$ ions in $Ni_3TeO_6$.** The exchange constants $J_1 - J_5$ are defined as follows: $J_1$ and $J_2$ are, respectively, between $Ni_I$ and $Ni_{II}$ of the edge-sharing $NiO_6$ octahedra in the *ab* plane and between $Ni_{II}$ and $Ni_{III}$ of the face-sharing $NiO_6$ octahedra along the *c* axis; both are ferromagnetic. $J_3$, $J_4$, and $J_5$ are between $Ni_{II}$ and $Ni_{III}$, $Ni_{III}$ and $Ni_I$, and $Ni_I$ and $Ni_{III}$ (corner-sharing $NiO_6$ octahedra), respectively, and all are antiferromagnetic. The difference between $J_4$ and $J_5$ derives from the variation in the $Ni_{III}$-O-$Ni_I$ bond distances and bond angles, where $Ni_I$ is below and above the $Ni_{III}$-Te honeycomb layer for $J_4$ and $J_5$, respectively.

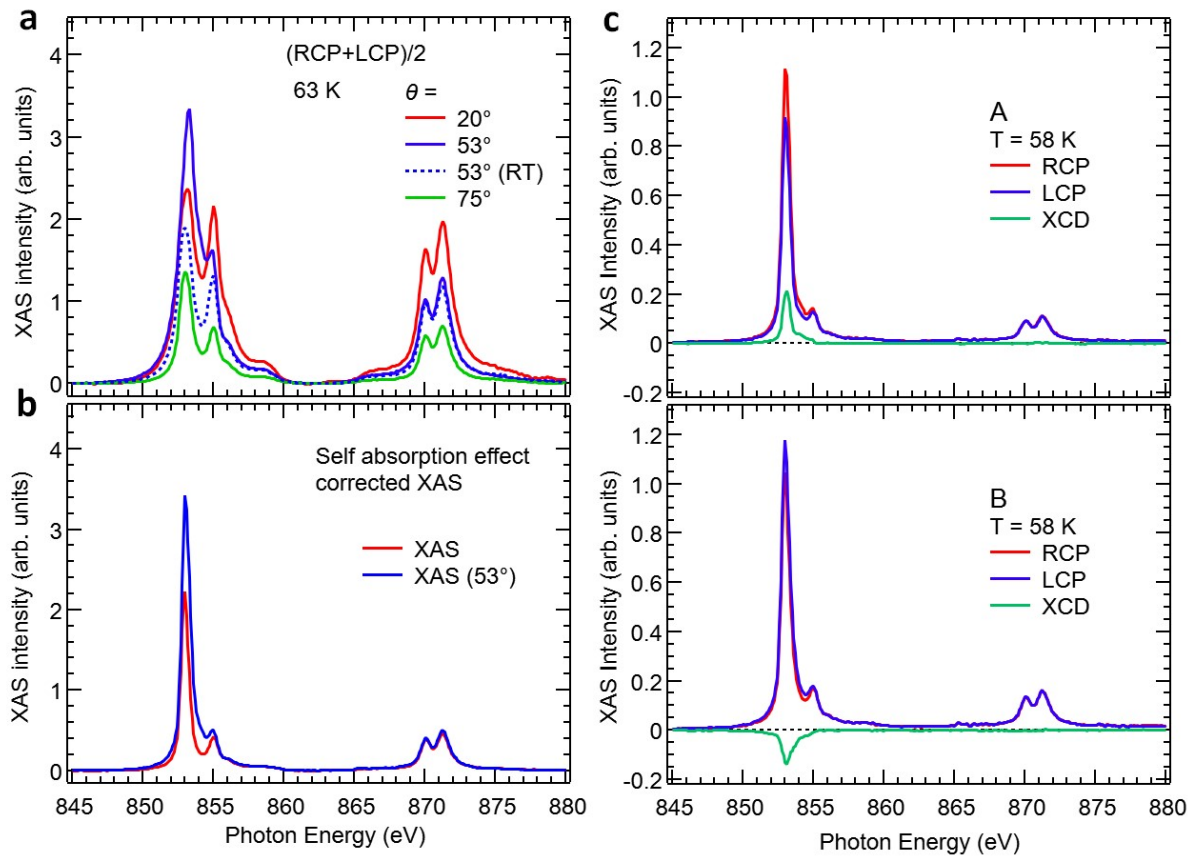

Figure S6: **Measured and self-absorption-corrected Ni $L_{2,3}$-edge XAS spectra of Ni$_3$TeO$_6$. a** Ni $L_{2,3}$-edge XAS (averaged over RCP and LCP and positions A and B) spectra of Ni$_3$TeO$_6$ measured at 63 K and the incident angles of $\theta$ = 20, 53, and 75°. Broken blue curve is XAS measured at $\theta$ = 53° and RT. **b** Self-absorption-corrected XAS spectra. The red curve is obtained from the measured XAS spectra of $\theta$ = 20 and 75°. The blue curve is obtained from the measured XAS spectrum of $\theta$ = 53° at 63 K multiplying a correction factor estimated from the XAS spectra of $\theta$ = 53° (RT). **c** Self-absorption corrected Ni $L_{2,3}$-edge XAS spectra at positions A and B with RCP and LCP and difference spectra (XCD) of $\theta$ = 53° at 58 K.

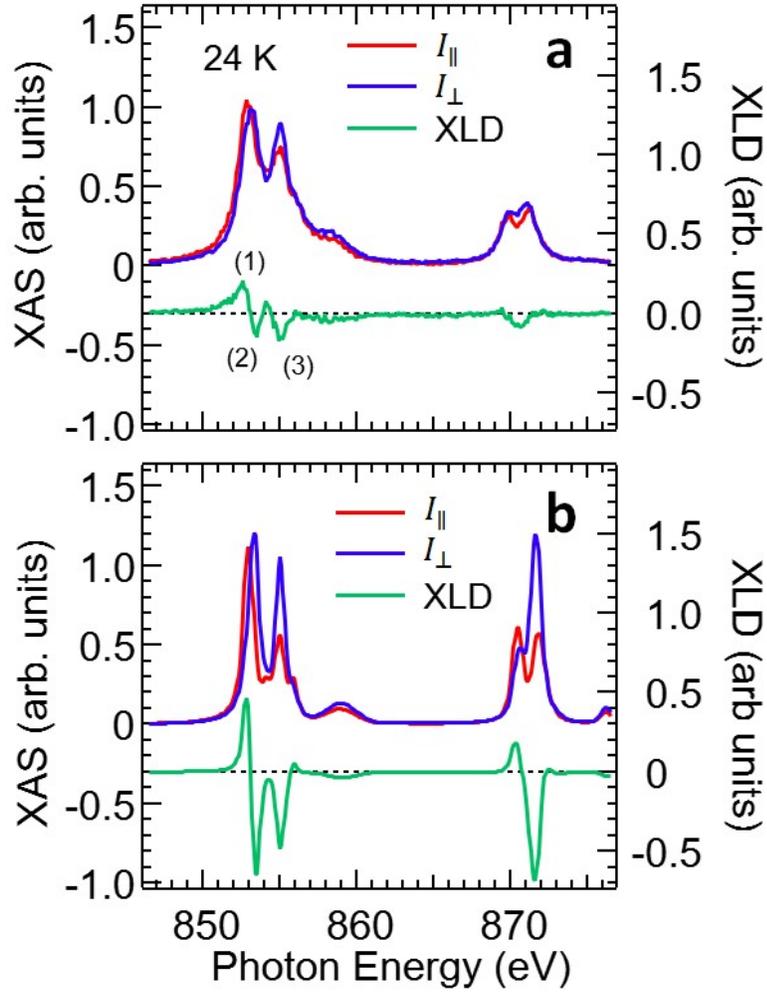

Figure S7: **XAS and XLD spectra of $Ni_3TeO_6$ compared with calculations. a**, Ni $L$-edge XAS spectra of $Ni_3TeO_6$ of incident X-rays with electric-field parallel ($I_∥$) and perpendicular to the $c$ axis ($I_⊥$) and their difference at 24 K. For the $I_∥$ measurement, the $E$ vector of the incident X-ray was in the $ac$ plane and 20° off the $c$-axis. The $I_∥$ spectra were deduced from the relation $I_∥ = \frac{1}{\cos^2\theta}(I_\pi - \sin^2\theta I_⊥)$, where $I_\pi$ is the measured spectra. **b**, XLD spectra calculated by using the $NiO_6$ cluster model with $C_{3v}$ symmetry. The crystal-field parameters were set to $10Dq$ = 0.9 eV, $D_\sigma$ = 0.05 eV, and $D_\tau$ = −0.03 eV. Charge-transfer energy $\Delta$ = 3.5 eV and Coulomb energies $U_{dd}$ = 7.2 eV and $U_{pd}$ = 8.0 eV were used for electron hopping between $d^8$ and $d^9\underline{L}$ configurations. The transfer integrals within $e^{(e_g)}$, $a_1^{(t_{2g})}$, and $e^{(t_{2g})}$ orbitals were set to 2.025, 1.55, and 1.65 eV for right-handed chirality and spin-up $Ni^{2+}$ sites, respectively.

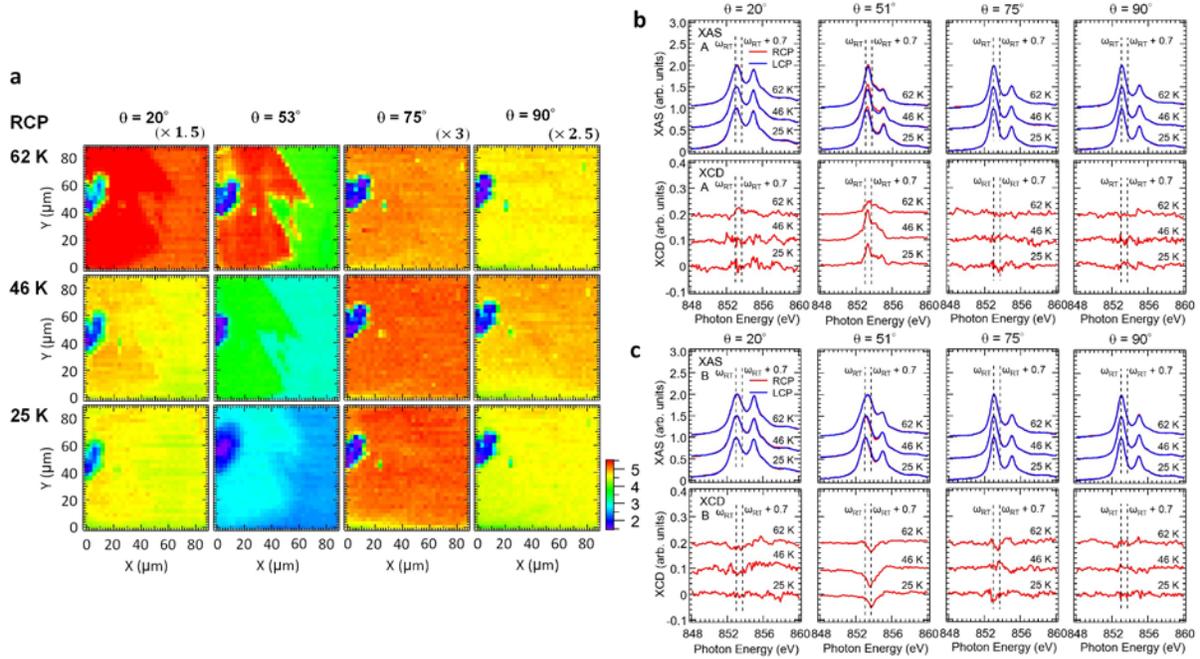

Figure S8: **Angle-dependent XAS images and XCD spectra in Ni$_3$TeO$_6$. a** XAS images in the 90×90-$\mu m^2$ area shown in Fig. 4**a** measured at the incident photon energy of $\omega_{RT}$+0.3 eV with RCP at various temperatures and the incident angles $\theta$ = 20,53,75, and 90°. For $\theta$ = 90°, the CCD detector was set at 105° off the incident X-rays. XAS intensities are shown by color and the parenthetical numbers show the magnification factors at each incident angle. XAS images at $\theta$ = 53° show clear a contrast between domains of opposite chirality. **b, c** Ni $L_3$-edge XAS spectra with RCP and LCP, and difference spectra XCD measured at the incident angles $\theta$ = 20,51,75, and 90° and various temperatures for domains A and B, respectively. XAS and XCD intensities are normalized to the $L_3$ average XAS peak intensity, (RCP+LCP)/2. The difference spectra of A and B at $\theta$ = 51° show strong positive and negative XCD structures.